\documentstyle[epsf,twocolumn,aps]{revtex}
\begin{document}

\twocolumn[\hsize\textwidth\columnwidth\hsize\csname @twocolumnfalse\endcsname

\title{Short-coherence length superconductivity in the Attractive Hubbard 
Model in three dimensions}

\author{Jan R.~Engelbrecht and Hongbo Zhao}
\address{Department of Physics, Boston College, Chestnut Hill, MA 02467}
\maketitle

\begin{abstract}

We study the normal state and the superconducting transition in the 
Attractive Hubbard Model in three dimensions, using self-consistent 
diagrammatics.
Our results for the self-consistent $T$-matrix approximation are consistent
with 3D-$XY$ power-law critical scaling and finite-size scaling.
This is in contrast to the exponential 2D-$XY$ scaling the method was
able to capture in our previous 2D calculation.
We find the 3D transition temperature 
at quarter-filling and $U=-4t$ to be
$T_c=0.207t$.
The 3D critical regime is much narrower than in 2D and the ratio of the
mean-field transition to $T_c$ is about 5 times smaller than in 2D.
We also find that, for the parameters we consider, 
the pseudogap regime in 3D (as in 2D) coincides with the
critical scaling regime.
\end{abstract}


\vskip1pc
]

\vspace{12pt}

\narrowtext

Inspired by the discovery of high-temperature
superconductors there has been considerable interest for
some time in understanding how the BCS picture 
gets modified in the case of short coherence length superconductivity.
A simple model for a correlated superconductor is the
Attractive Hubbard Model that allows one to tune the attraction
to an intermediate regime  between the BCS and Bose limits
\cite{3D-a} where the fermions are still degenerate but the coherence
length starts to approach the interparticle spacing.
Earlier studies \cite{NEGU_NS1} of this model in 2D using Monte Carlo 
methods revealed the development of a pseudogap in the normal state.
In this paper we employ self-consistent diagrammatics
that is a variant of the Fluctuation-Exchange Approximation 
\cite{FLEX1,FLEX2,FLEX3,FLEX4}.
Specifically, we use the self-consistent $T$-matrix approximation (STA)
that only retains fluctuations in the particle-particle channel.
These are presumably the only relevant fluctuations in the model for the
densities we consider.
%
%
We have found before \cite{EN} that quite remarkably, the STA approach in 2D
captures the critical 2D-$XY$ fluctuations with a Kosterlitz-Thouless (KT)
transition temperature in good agreement with 
Monte Carlo \cite{Moreo_Scalapino}.  In addition
we found a very broad critical regime that coincides with the
development of a normal state pseudogap.
It is very surprising that self-consistent diagrammatics can capture
critical behaviour relatively well.
In order to understand this better, and encouraged by the success in 2D,
we now study this model in three dimensions where techniques like
Monte Carlo are severely limited by system size constraints.

We find that the STA method indeed yields results in three dimensions
that are consistent with the expected power-law scaling rather than the
exponential KT-scaling we found in 2D.  In addition, the success of
finite-size scaling allows us to determine $T_c$ in 3D.  
In addition to observing a much narrower pseudogap regime than in 2D,
we can quantify the difference in pairbreaking fluctuations by looking at the
ratio of the mean-field (1-loop) $T_c$ to the actual
transition temperature and find that this ratio drops from 12 to about 2.6
in going from 2D to 3D, for the parameters we consider.
We also conclude that since the dramatic KT-scaling observed in 2D
is not seen in 3D, this result is not a numerical artifact.

While superconductivity in the Attractive Hubbard Model is of course
$s$-wave, it is fair to assume that the strong correlation effects in
this model will also apply to a more realistic $d$-wave theory.  In fact,
an advantage of the STA is that one can explicitly consider models with
$d$-wave attraction\cite{ENRD}.
In the context of the cuprates our work suggests that the loss of low-energy
spectral weight in the pseudogap normal state can be attributed to
pre-cursor pairing as observed in our calculation.
By studying dimensionality, the observation then is that
underdoped cuprates behave more like our 2D calculations while optimally
and overdoped materials have pseudogap behaviour more in line with our 3D
results.  Since, within a given family of materials, the anisotropy can
increase several hundred-fold upon underdoping, the implication is that
this change in effective dimensionality plays an important role in enhancing
the underdoped pseudogap.

Our starting point is the hamiltonian for the Attractive Hubbard Model 
\begin{equation}
H=-t\sum_{\langle ij\rangle\sigma}
c^\dagger_{i\sigma}c_{j\sigma}
+hc
-|U|\sum_i {n}_{i\uparrow} {n}_{i\downarrow}
\end{equation}
where the tight-binding hopping sums over nearest neighbour pairs
on a three-dimensional square lattice
and the potential captures singlet pairing for a short-range potential.
The standard perturbative approach is to evaluate the
pair susceptibility at the 1-loop level.  Such a Gaussian theory amounts to
re-summing the BCS ladder diagrams yielding an RPA-like form 
\begin{equation}
\chi_{gaus}({\bf q},i\omega_n)={\Pi_0({\bf q},i\omega_n)\over
1-|U|\Pi_0({\bf q},i\omega_n)}
\label{eq-rpa}
\end{equation}
where $i\omega_n$ are Matsubara frequencies and $\Pi_0$ is the bare bubble
in terms of free fermion propagators.
In the limit of small $U$, this framework reduces to BCS-theory. 
We are interested in intermediate attraction where the
fermions are still degenerate but the pairing is strong.
The value $U=-4t$ that we use throughout this paper is
representative of such intermediate attraction.
The model also deviates strongly from BCS theory near half-filling,
where the Attractive Hubbard Model maps into the Hubbard Model with
on-site repulsion.  Since we do not wish to include the effects of the
associated CDW instability at half-filling, all our results in this
paper will focus on quarter-filling.

The superconducting instability signaled by the Thouless criterion
that determines the transition temperature as the highest
temperature that satisfies
${1/\chi_{gaus}(T=T_c)}=0$.

Our approach is to extend this framework to the self-consistent $T$-matrix 
(STA) formulation, which simply
generalises the bare bubble in (\ref{eq-rpa}) with dressed propagators
for the intermediate fermion states:
\begin{equation}
\Pi({\bf q},i\omega_n)=
{-1\over\beta}
\sum_{{\bf k},ik_n} 
G({\bf q}\!-\!{\bf k},i\omega_n\!-\!ik_n) G({\bf k},ik_n)
\label{eq-pi}
\end{equation}
where the dressed Green's function $G({\bf k},ik_n)=[ik_n-\epsilon_{\bf k}+\mu-
\Sigma({\bf k},ik_n)]^{-1}$ 
is defined in terms of the pair susceptibility itself, through the
self-energy:
\begin{equation}
\label{eq-se}
\Sigma({\bf k},ik_n)= 
{1\over\beta}
\sum_{{\bf q},i\omega_n}
G({\bf q}\!-\!{\bf k},i\omega_n\!-\!ik_n) \Gamma({\bf q},i\omega_n)
.
\end{equation}
Technically, we explicitly keep only corrections of order $U^2$ and
higher which implies that the Hartree term is already included in the
unperturbed chemical potential.  This higher order
contribution is expressed though the effective interaction:
\begin{equation}
\Gamma({\bf q},i\omega_n)={U^2\;\Pi({\bf q},i\omega_n)\over
1-|U|\Pi({\bf q},i\omega_n)}
.
\label{eq-Gamma}
\end{equation}

The self-consistent solution of equations
(\ref{eq-pi}) and (\ref{eq-se}) are achieved numerically, by considering
a finite {\it N}$\times${\it N}$\times${\it N}
lattice with discrete momenta and 
only retaining a finite number of Matsubara frequencies
\cite{foot-infinite-matsubara}.
We start off with the unperturbed solution and iteratively solve these
equations (which are convolutions) using FFT's until self-consistent 
convergence is established.

Perturbatively this of course amounts to re-summing an infinite series of
diagrams.  At first one may expect that the self-consistent feedback will 
broaden the intermediate fermion states in (\ref{eq-pi}) and
simply re-scale values for physical quantities resulting from a 1-loop 
calculation.  This is in fact not the
case as can be seen from the results in Fig.~1 for the inverse of the
uniform static pair susceptibility $\chi({q\!=\!0},i\omega_n\!=\!0)$ evaluated for
quarter-filling $\langle n_{i\sigma}\rangle=\frac14$ and the intermediate
attraction $U=-4t$.
The results are for a 32$\times$32$\times$32 lattice.
On the RHS the 1-loop result intersects the
axis at $T_c^{gaus}(3D)=0.532t$.
This result is essentially independent of lattice size
and the 1-loop approximation yields a non-zero transition temperature
even for finite systems.
In any exact treatment, one of course only expects phase transitions
in the thermodynamic limit.
We also indicate the 2D transition temperature in this 1-loop approximation
for the same parameters at $T_c^{gaus}(2D)=0.611t$.

\begin{figure}
\vspace{-20pt}
  {\epsfxsize=3.5truein\epsfbox{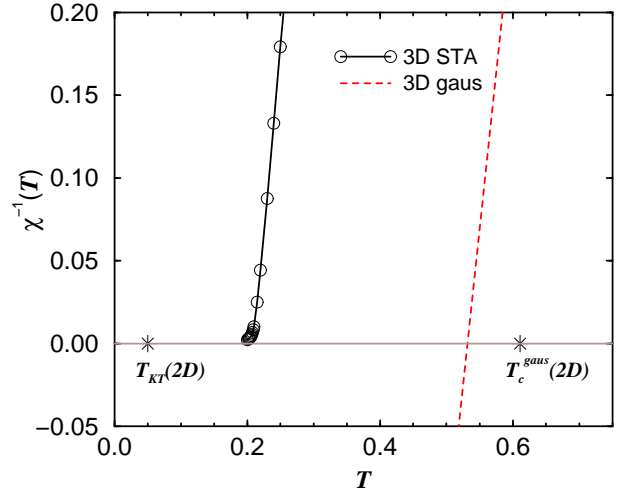}}
\vspace{-6pt}
\caption{
Inverse pair susceptibility for a 
3D lattice of linear length $L=32$, 
density $\langle n_{i\sigma}\rangle=\frac14$ and 
$U=-4t$.
The symbols indicate the STA results and the 
broken line on the right is for the 1-loop (mean-field) theory.  
Also shown are the 2D mean-field and KT transition temperatures.
}
\end{figure}

The 3D STA results indicated by the circles 
show a fundamentally different behaviour.  The results for the inverse
susceptibility never cross zero -- indicating that the 
self-consistent calculation reflects the fact that there is no transition
for a  finite system.  This is of course what one would expect from an
exact treatment - but is missed by 1-loop theory.  
Simply put, the self-energy corrections broaden the intermediate fermion
propagators sufficiently to avoid the superconducting instability.
Thus the STA is fundamentally superiour to 1-loop in that it captures
the physical requirement that there are no phase transition in finite systems.

\begin{figure}
\vspace{-20pt}
  {\epsfxsize=3.5truein\epsfbox{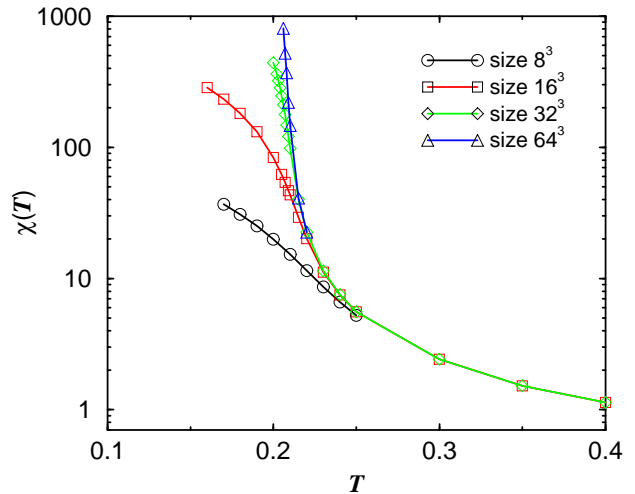}}
\vspace{-6pt}
\caption{
Finite-size dependence of $\chi(T,L)$ in 3D for
$\langle n_{i\sigma}\rangle=\frac14$ and $U=-4t$.
}
\end{figure}

Is there any more to this result?
Further investigation reveals that the low-temperature 
static uniform pair susceptibility $\chi(T,L)$ for the STA
depends on system size as shown in Fig.~2.  
Just as one would expect from fundamental principles,
the susceptibility increases towards a maximum that is controlled by
system size and can only diverge in the thermodynamic limit.
Since the formalism does not include anomalous propagators, it is
limited to the disordered side of the phase diagram.
This plot is quite similar to our earlier results\cite{EN} in 2D, 
except that the attempt at
a transition appears sharper and is at a higher temperature in 3D.

From this finite-size data we use standard scaling arguments
\cite{textbook}
to extract the transition temperature in 3D.
For a finite system there are two length scales namely the
correlation length $\xi$ and the system size $L$.
One expects from a 3D scaling hypothesis that
$\displaystyle
\chi(T,L)=\left|T-T_c\right|^{-\gamma}f_1(L/\xi(T))
$
where $\xi(T)$ is the correlation length in an infinite system.
From $\xi(T)\sim\left|T-T_c\right|^{-\nu}$ and the scaling law
$\gamma=\nu(2-\eta)$, it follows that the following product is only a function
of the ratio of the two length scales:
\begin{equation}
\label{eq-chtl2}
\left({1\over L}\right)^{2-\eta}\chi(T,L)=f(L/\xi(T))
.
\end{equation}

At $T_c$ the correlation length diverges and the argument on the RHS
becomes zero, independent of system size.  Plots for this function for
our STA data on different system sizes are given in Fig.~3.
The numerical data is consistent with the finite-size scaling hypothesis
and we can estimate $T_c=0.207t$ for our standard set of parameters.
This value of $T$ is indicated by the vertical line where the
plots become independent of $L$.

\begin{figure}
\vspace{-20pt}
  {\epsfxsize=3.5truein\epsfbox{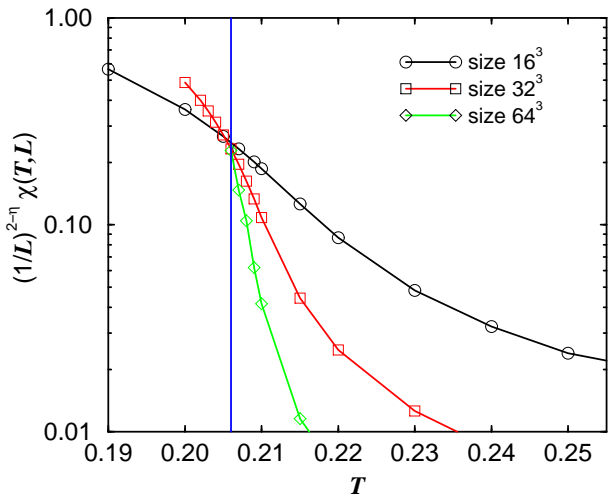}}
\vskip 0.0truecm
\caption{
Finite-size scaling of $f(L/\xi(T))$ in 3D for
$\langle n_{i\sigma}\rangle=\frac14$ and $U=-4t$.
The common intersection yields $T_c=0.207t$.
}
\end{figure}

In our earlier 2D calculation, the STA yielded results that were very 
consistent with the Kosterlitz-Thouless scaling of the 2D-$XY$
model, with an estimate of $T_{KT}=0.05t$ in good agreement with
previous Monte Carlo estimates\cite{Moreo_Scalapino}.  
In addition, we found very broad
critical scaling regime that coincides with the regime in which a 
pseudogap develops in the density of states at the Fermi level,
$N_0(T)$.  We also found very good finite-size scaling in our data.
Given the difficulty of theoretically capturing the correct critical
behaviour this was very surprising but one could always question whether
this result was a numerical artifact.

Here we have demonstrated, that in 3D the STA data is again consistent with 
a finite-size scaling hypothesis -- yielding the first estimate of 
$T_c=0.207t$ for the parameters in question, in a calculation that goes beyond 
the mean-field or 1-loop class of approaches.

It is interesting to note that while the 1-loop estimate for the
transition temperature is higher in 2D than in 3D for the same parameters
(due to density of states effects) the opposite holds for the actual
transition temperatures.  We conclude that the enhanced 
fluctuations in 2D overcome the density of states effect.
If we consider the ratio of mean-field to actual transition temperatures
$R=T_c^{gaus}/T_c$ we find $R_{2D}=0.611/0.05\sim12$ {\it vs.}
$R_{3D}=0.532/0.207\sim2.6$.

Let us next consider the bulk scaling behaviour.
Is this consistent with the expected 3D-$XY$ result
\begin{equation}
\chi(T)\sim{1\over\left|T-T_c\right|^{\gamma}}
\label{eq-powerlaw}
\end{equation}
with $\gamma\simeq1.33$?
In 3D we are limited to smaller systems than in 2D and the transition
itself is narrower.  This limits the accuracy with which we can 
determine the critical scaling in the thermodynamic limit within our
calculations.  In Fig.~4 we plot the finite-size data $\chi(T,L)$
against 
the RHS of (\ref{eq-powerlaw}).  
\begin{figure}
\vspace{-20pt}
  {\epsfxsize=3.5truein\epsfbox{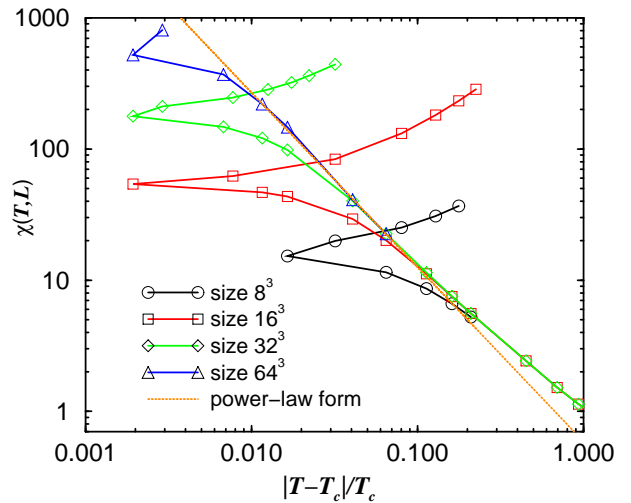}}
\vskip 0.0truecm
\caption{Finite-size results for $\chi(T,L)$ and the expected scaling
form in (\ref{eq-powerlaw}) with exponent $\gamma\simeq1.33$.
}
\end{figure}
\noindent
For each system size, the
high-temperature data coincides  with the scaling
form in (\ref{eq-powerlaw}) and at lower temperature one observes the
expected finite-size corrections.  While the data is clearly consistent 
with power-law scaling and the fit with the exponent $\gamma\simeq1.33$
does better than the mean-field $\gamma_{MF}=1$, our accuracy is insufficient
to determine the exponent a-priori.
In addition we can estimate the width of the critical regime as extending
about 0.1$t$ above $T_c$, where the high-temperature results
start to deviate from the asymptotic form.

While we cannot determine the power-law exponent, the 3D data is
dramatically different from our previous 2D result that found
data consistent with the exponential KT scaling expected in 2D.  
This gives some confidence that the 2D result was real
and that the STA method captures the expected critical behavour
in both 2D and 3D quite well.
This implies that for this particular model, self-consistency 
introduces scaling corrections that go beyond dimensional analysis
-- a feat that 
normally requires a technique such as the renormalization group.
We believe that this may be connected to the rather simple $U(1)$
symmetry that superconductivity breaks, and that this result 
does not apply to more complicated problems.

Let us finally turn to the normal state pseudogap.
The density of single-electron states can be obtained
from the Green's functions dressed with our calculated
self-energy corrections $\Sigma({\bf k},ik_n)$ and 
continued to retarded frequencies using
Pad\`e analytic continuation.
In Fig.~5 we show the temperature-dependence of the density of states
at the Fermi level for our standard density and interaction strength.
We show results for both 3D and 2D and indicate the respective $T_c$'s
by the vertical lines.
One can clearly see that the 3D pseudogap is significantly smaller
than in 2D and again coincides with the critical regime.
In earlier analytic work \cite{3D-E}, we calculated the 3D pseudogap
for a continuum model within the 1-loop theory
and found that even at this level, the pseudogap corrections
go beyond the expected results in a small energy, long-wavelength
expansion.

\begin{figure}
\vspace{-20pt}
  {\epsfxsize=3.5truein\epsfbox{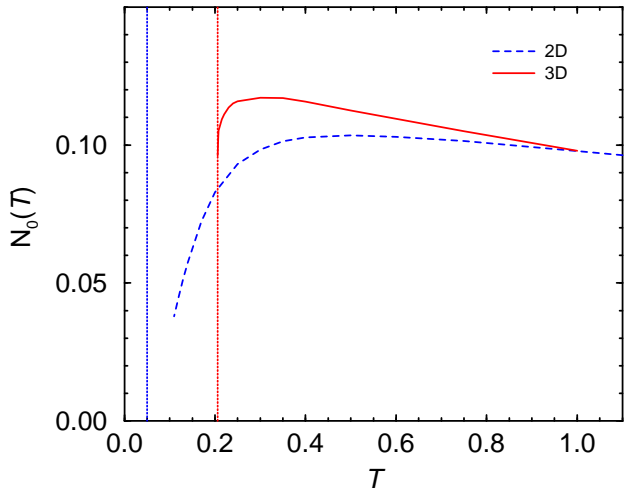}}
\vskip 0.0truecm
\caption{Pseudogap in the  density of states at at the Fermi level 
in 3D and in 2D 
for $\langle n_{i\sigma}\rangle=\frac14$ and $U=-4t$.
}
\end{figure}

In summary, we have shown that results for the self-consistent $T$-matrix 
approximation in 3D are consistent with the expected
finite-size and power-law critical behaviour for the 3D-$XY$ model.
At quarter-filling and for and intermediate attraction $U=-4t$,
we find in 3D $T_c\simeq0.207t$, about 2.6 times smaller than the mean-field
transition temperature, and a narrow critical and pseudogap regime.
We also find that while the mean-field transition temperature is
higher in 2D than in 3D, the opposite holds for the actual $T_c$.
Our results suggest that the previous 2D results are not a numerical
artifact and that the STA formalism can distinguish between the
fundamentally different nature of fluctuations in two and three dimensions.

We wish to acknowledge useful discussions with Alexander Nazarenko who
contributed in a key way to our earlier work using the STA technique.





\begin{references}
\bibitem{foot-infinite-matsubara}
Technically, we incorporate contributions from higher-order
Matsubara frequencies analytically, wherever possible.
\end{references}
\end{document}